\documentclass[aps, prd, 10pt, twocolumn, nofootinbib,floatfix,superscriptaddress]{revtex4-2}
\pdfoutput=1
\usepackage{subfigure}
\usepackage{epsfig}
\usepackage{amsmath}
\usepackage{amsfonts}
\usepackage{amssymb}
\usepackage{booktabs}
\usepackage{siunitx}
\usepackage{xfrac}
\usepackage{color}
\usepackage{cancel}
\usepackage[utf8]{inputenc}
\usepackage{graphicx}
\usepackage{dcolumn}
\usepackage{bm}
\usepackage{tikz}
\usepackage{ragged2e}
\usepackage{tabularx}

\usepackage{tcolorbox}
\usepackage{hyperref}
\hypersetup{colorlinks, citecolor=red, linkcolor=bluscuro, urlcolor=bluscuro}
\definecolor{rossos}{cmyk}{0,1,1,0.55}
\definecolor{bluscuro}{rgb}{0.15, 0.2, .85}
\definecolor{bluchiaro}{cmyk}{1,.3,0.,0.1}

\makeatletter
\long\def\@makecaption#1#2{%
  \vskip\abovecaptionskip
  \sbox\@tempboxa{#1: #2}%
  \ifdim \wd\@tempboxa > \linewidth
    \begin{center}
      \parbox{\linewidth}{\justifying #1: #2}
    \end{center}
  \else
    \global\@minipagefalse
    \hb@xt@\linewidth{\hskip 0pt #1: #2\hfil}%
  \fi
  \vskip\belowcaptionskip}
\makeatother

\begin{document}

\preprint{}

\title{A Reproducible Inertia Dynamometer for Vocational Schools:
   Design, Fabrication, and Pedagogy}

\author{Stylianos A.\ Tsilioukas}
\email{tsilioukas@sch.gr}
\affiliation{\mbox{National Observatory of Athens, Lofos Nymfon, 11852 Athens, 
Greece}}

\begin{abstract}
We report the design, construction, and classroom use of an inertia
dynamometer, built as a year-long project-based learning (PBL) activity with
adult students at a Greek Evening Vocational High School (EPAL). The apparatus
consists of a machined steel drum of calculated moment of inertia
$I = 0.6507~\mathrm{kg\,m^2}$, mounted on a student-welded frame and
instrumented with a green-laser / light-dependent resistor (LDR) optical
interrupter. The analogue output is sampled at 44.1\,kHz by the microphone
input of a laptop computer, which is used as an opportunistic analogue-to-digital
converter; torque and power curves are then reconstructed in software from
the inter-pulse intervals via $\tau = I\alpha$ and $P = \tau\omega$. The
drum's moment of inertia is cross-checked by an inclined-plane rolling
experiment. A wide-open-throttle test with a 50\,cc scooter reproduces the
expected flat-power / falling-torque signature of a continuously variable
transmission in the low-to-moderate RPM range; the LDR's millisecond-scale
recovery time imposes an upper bandwidth limit that provides an unplanned
but pedagogically rich lesson in sensor physics. The project integrated
industrial-lathe fabrication, arc welding, analogue electronics, and
numerical differentiation into a single coherent workflow. We describe the
apparatus, the physics, the signal-processing pipeline (for which MATLAB
and Python/Octave code are provided as supplementary material), and reflect
on the pedagogical outcomes for a student population traditionally
disengaged from abstract physics.
\end{abstract}

\maketitle

\section{Introduction}
\label{sec:intro}

Rotational kinematics, the interplay of torque, moment of inertia, and
angular acceleration, remains a well-documented source of difficulty in
introductory
physics~\cite{Mashood_2015, PhysRevSTPER.1.010102, close2011angular}.
Students frequently fail to distinguish angular and translational
quantities, or to apply Newton's second law for rotation correctly,
even after formal instruction~\cite{PhysRevSTPER.1.010102}.
The difficulty is amplified in Vocational Education and Training (VET)
environments, where the learner population differs markedly from that of
a typical secondary school. In the Greek Evening Vocational High Schools
(\textit{EPAL}), students are predominantly adult workers returning to
formal education after extended interruptions; they arrive at evening
classes already fatigued by a day of manual labour, and many have
histories of disengagement from traditional, lecture-driven instruction.

These students nonetheless bring a substantial body of informal
mechanical knowledge, particularly around motorcycles and small engines,
which are culturally prominent in the region. They have tacit intuitions
about power, torque, and mechanical resistance, without the formal
vocabulary to describe them. Research on contextualised instruction
shows that anchoring abstract content in students' prior knowledge and
real-world experience improves both engagement and
understanding~\cite{https://doi.org/10.1002/tea.20203}. More broadly,
project-based learning (PBL) has been shown to increase motivation and
problem-solving ability in vocational
cohorts~\cite{chiang2016effect,educsci15080941}. The rise of
open-source hardware and commodity electronics has, at the same time,
made it possible to build laboratory-grade apparatus at school
budgets~\cite{Bouquet_2017,pearce2017impacts,dausilio2012arduino}.

This paper reports a year-long PBL project in which a class of adult
EPAL students designed, fabricated, instrumented, and calibrated an
inertia dynamometer capable of producing a torque--speed curve for a
small motorcycle. An inertia dynamometer measures engine output by
accelerating a flywheel of known moment of inertia; the angular
acceleration of the flywheel, combined with Newton's second law for
rotation ($\tau = I\alpha$), yields the instantaneous
torque~\cite{martyr2012engine,Buschert2003}. It is worth distinguishing this instrument from the object most physics
students associate with the word ``dynamometer'': the spring based force
gauge governed by Hooke's law, $F = kx$, which reports a single static
force from a linear displacement. A torsion balance is the rotational
analogue of that static instrument; it infers a torque from the angular
deflection of a calibrated spring or fibre, again under equilibrium
conditions. An inertia dynamometer differs from both in kind, not only
in degree: it produces no static reading at all. Instead, it exploits
Newton's second law for rotation under active acceleration,
$\tau = I\alpha$, inferring the driving torque from how quickly a
flywheel of known moment of inertia is spun up. The measurement
therefore exists only while the system is accelerating, and its accuracy
depends on differentiating a kinematic signal rather than reading an
equilibrium deflection, a distinction we return to in Section~V.
Our apparatus is deliberately
minimalist: a heavy steel drum of known~$I$, a photogate made from a
laser pointer and a cadmium-sulphide photoresistor, and the microphone
input of a laptop computer used as a makeshift analogue-to-digital
converter, an approach that has been validated for high-resolution
timing in other physics
contexts~\cite{Aguiar_2010,hassan2011soundcard}.

Our aims are threefold: to describe a complete, reproducible, low-budget
apparatus; to document the signal-processing pipeline and its numerical
pitfalls; and to reflect on the pedagogical outcomes for an adult
vocational cohort, including an instructive sensor-bandwidth failure that
became one of the most productive teaching moments of the year.
The project was carried out in 2010; the present account is
retrospective and does not include a formal assessment of learning
gains, so we frame the pedagogical observations as a structured case
study.

\section{Mechanical Construction}
\label{sec:construction}

The mechanical build was treated as a core pedagogical exercise:
grounding the abstract quantity $I$ in a piece of steel the students
could lift, and connecting the classroom to local craftsmen and
scientific institutions.

\subsection{The inertia drum}

\begin{figure}[htbp]
    \centering
    \includegraphics[width=\columnwidth]{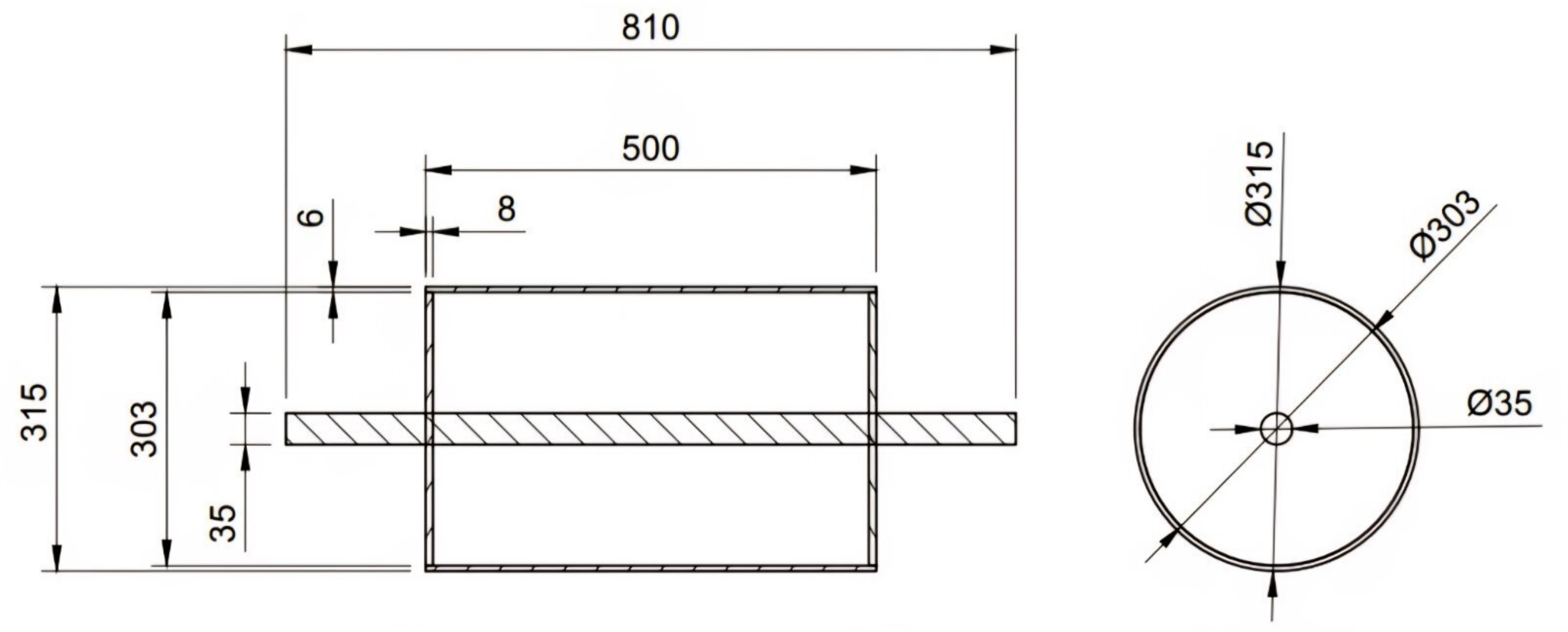}
    \caption{Engineering drawing of the drum and shaft assembly,
    exported from the Fusion~360 CAD model that can be accessed in the repository \cite{Tsilioukas:2026repo}.}
    \label{fig:2D_drum_dim}
\end{figure}

The rotating assembly consists of a solid steel shaft, two internal
mounting disks, and a heavy outer cylindrical pipe (Fig.~\ref{fig:2D_drum_dim}), all fabricated from standard mild steel and welded together
(dimensions and moment-of-inertia contributions in
Section~\ref{sec:inertia}). The students drafted the geometry; final
machining was carried out at a local workshop, where the assembly was
turned on an industrial lathe to a concentric cylinder and dynamically
balanced. Filming the process provided a direct visual link between the
symbolic $I$ on the blackboard and a 38\,kg object rotating at
industrial speeds.

\subsection{The frame}

\begin{figure}[htbp]
    \centering
    \includegraphics[width=\columnwidth]{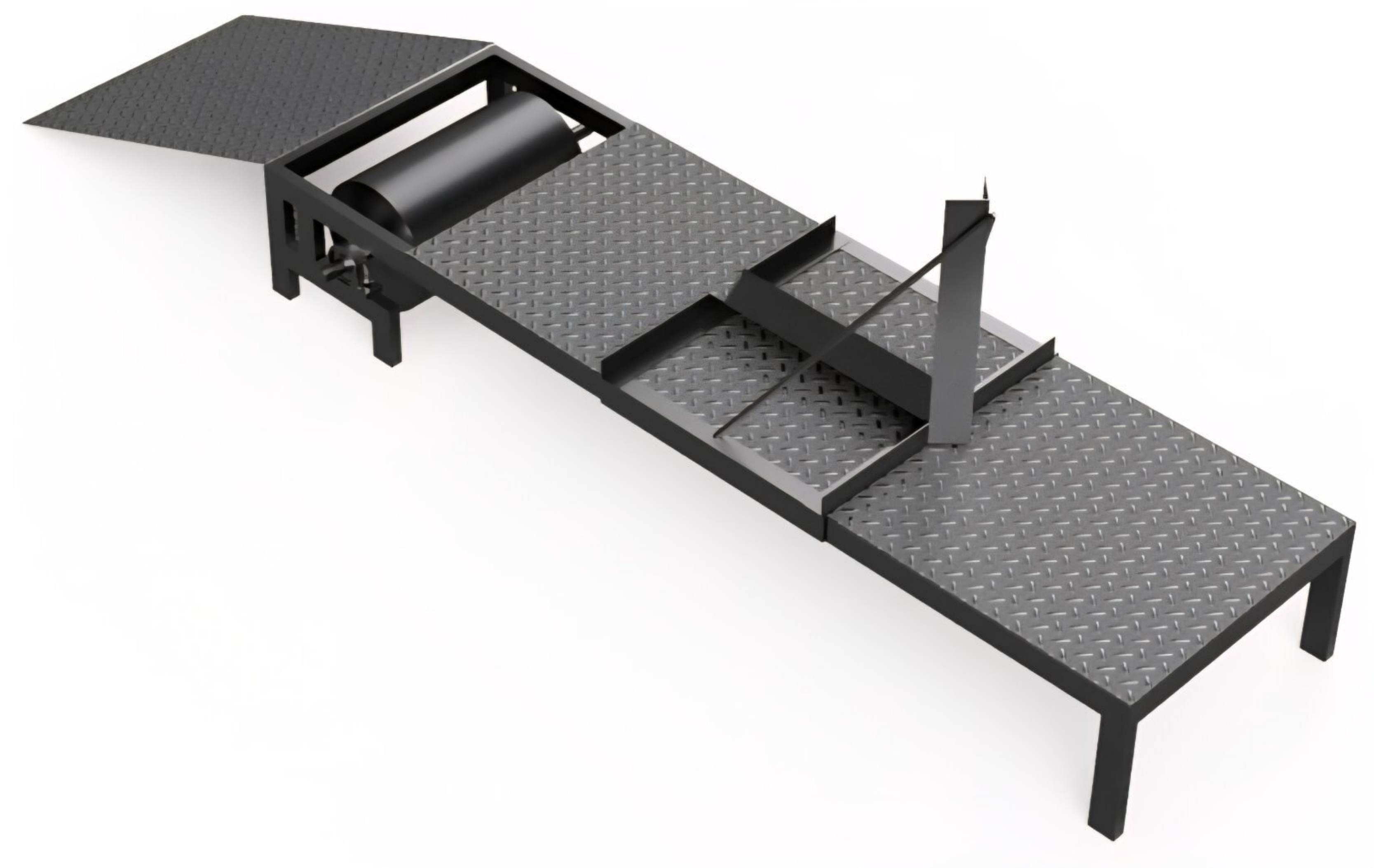}
    \caption{CAD rendering of the full dynamometer assembly
    (Fusion~360). The model, together with all part files, is
    available in the supplementary repository \cite{Tsilioukas:2026repo}.}
    \label{fig:3D_dyno}
\end{figure}

\begin{figure}[htbp]
    \centering
    \includegraphics[width=\columnwidth]{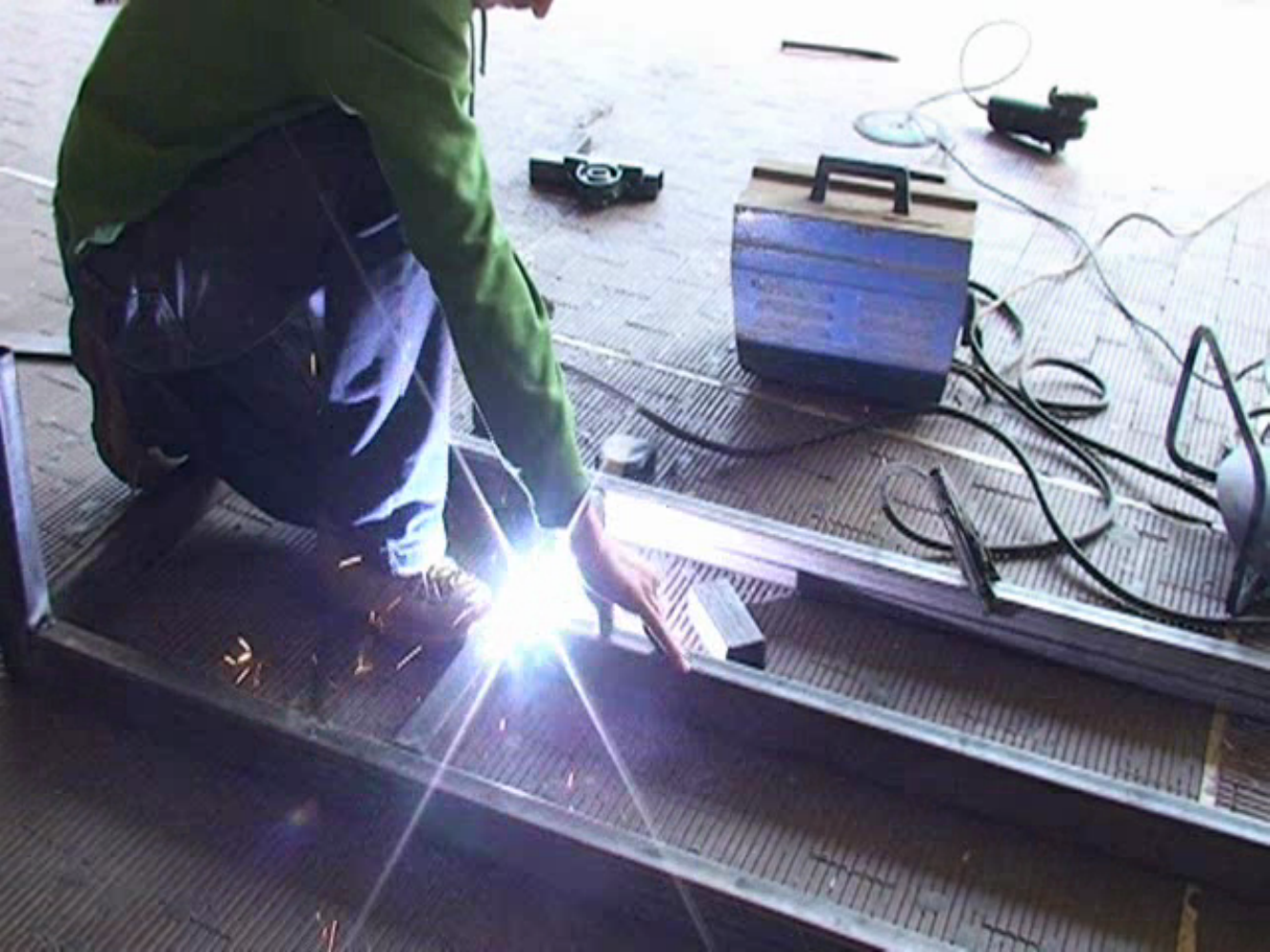}
    \caption{Students arc-welding the steel frame in the schoolyard.
    Fabrication by the students themselves was both a practical
    necessity (the project had no equipment budget) and a
    pedagogical choice intended to validate existing vocational
    skills.}
    \label{fig:frame_welding}
\end{figure}

The structural frame supporting the drum, its bearings, and the
motorcycle wheel-chock rollers was designed
(Fig.~\ref{fig:3D_dyno}) and built entirely by the students in the
school yard, using arc welding (Fig.~\ref{fig:frame_welding}).

This phase produced a notable and unplanned role reversal: the
students, experienced in practical trades, instructed the physics
teacher in welding technique, while the teacher contributed the
mechanical-design calculations. The inversion of expert/novice roles
was remarked on by the students themselves and appeared to contribute
positively to their sense of self-efficacy~\cite{bandura1977self}
and ownership of the project.

\subsection{Safety considerations}

A 38\,kg drum at several hundred rad\,s$^{-1}$ stores substantial
kinetic energy and demands corresponding precautions: dynamic
balancing by the professional workshop; a steel guard around the drum
perimeter and mandatory eye protection during tests. Arc welding was supervised by qualified
staff with appropriate Personal Protective Equipment (PPE). The Class~2--3R green laser was enclosed so
the beam was never exposed during operation. Any replication must
reproduce at least these precautions.

\section{Determining the Moment of Inertia}
\label{sec:inertia}

Because the reconstructed torque is strictly proportional to $I$
($\tau = I\alpha$), any systematic error in $I$ propagates directly
into the final horsepower figure \cite{Eadkhong_2012}. We therefore estimated $I$ in two
independent ways.

\subsection{Geometric calculation}

Using mild-steel density $\rho = 7850\,\mathrm{kg\,m^{-3}}$, the dimensions as are shown in Fig.~\ref{fig:2D_drum_dim} and the
standard rigid-body
formulae~\cite{halliday2013fundamentals}, the three rotating
components contribute as follows. For the solid central shaft
($R = 17.5$\,mm, $L = 810$\,mm):
\begin{equation}
    I_\mathrm{shaft} = \tfrac{1}{2} m R^2
                     = 9.36 \times 10^{-4}~\mathrm{kg\,m^2}.
\end{equation}
For the heavy outer pipe ($R_\mathrm{out} = 157.5$\,mm,
$R_\mathrm{in} = 151.5$\,mm, $L = 500$\,mm):
\begin{equation}
    I_\mathrm{pipe} = \tfrac{1}{2} m_\mathrm{pipe}
                       \bigl(R_\mathrm{out}^{2}+R_\mathrm{in}^{2}\bigr)
                   = 0.5458~\mathrm{kg\,m^2}.
\end{equation}
For the two internal mounting disks, each of mass $m_\mathrm{d}$
($R_\mathrm{out} = 151.5$\,mm, $R_\mathrm{in} = 17.5$\,mm, thickness
$t = 8$\,mm):
\begin{equation}
    I_\mathrm{disks} = 2 \times \tfrac{1}{2} m_\mathrm{d}
                       \bigl(R_\mathrm{out}^{2}+R_\mathrm{in}^{2}\bigr)
                    = 0.1039~\mathrm{kg\,m^2}.
\end{equation}
Summing, and cross-checking the total mass $m_\mathrm{tot}
= 38.0$\,kg against a direct scale measurement of the drum, we obtain
\begin{equation}
    I_\mathrm{theory} = 0.6507~\mathrm{kg\,m^2}.
\end{equation}

\subsection{Inclined-plane cross-check}

\begin{figure}[htbp]
    \centering
    \includegraphics[width=\columnwidth]{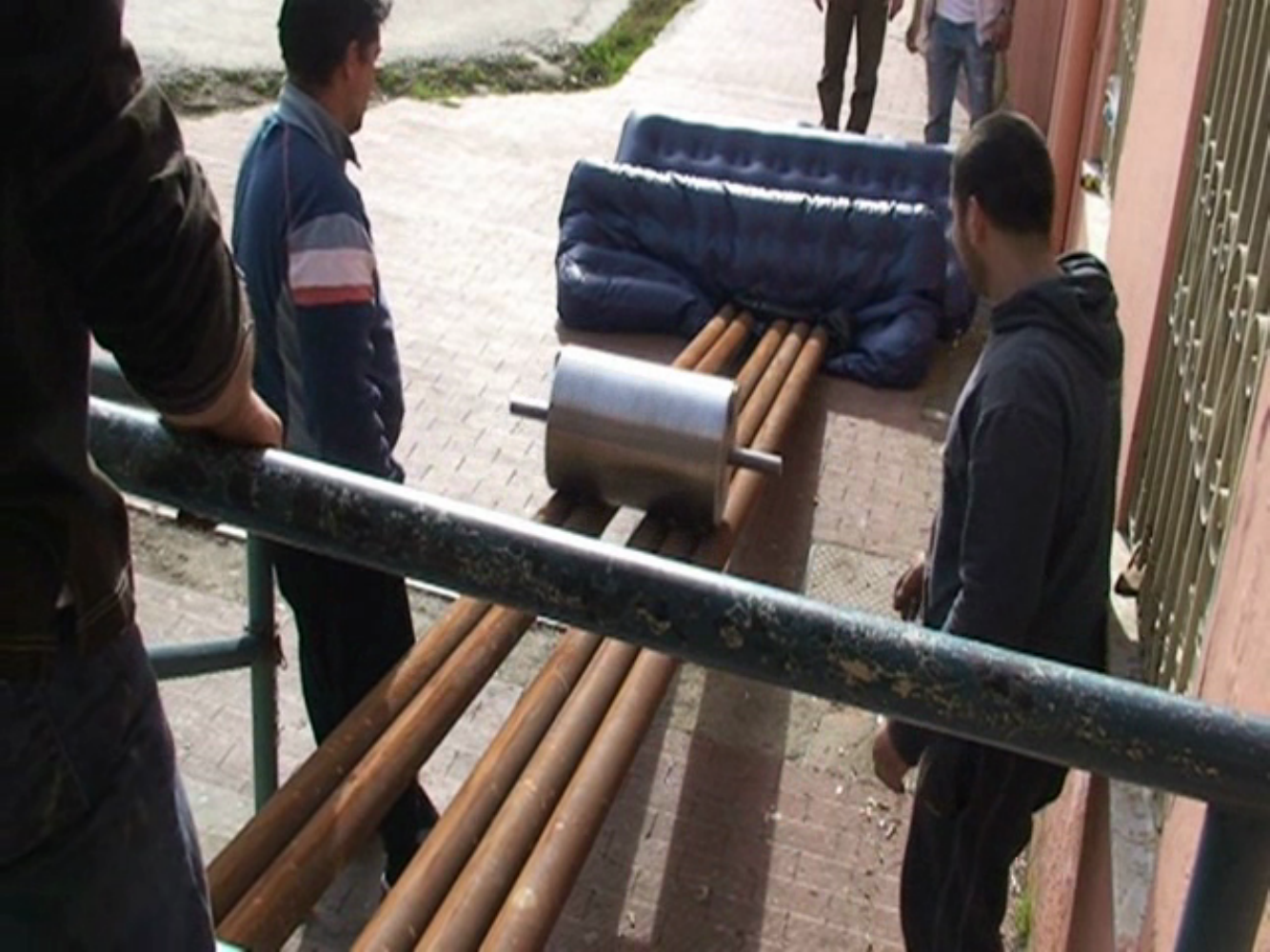}
    \caption{Inclined-plane cross-check of the drum's moment of
    inertia. The 38\,kg assembly rolls on four parallel steel rails
    inclined at an angle $\theta$; distance travelled and descent
    time are extracted from 100\,fps video footage.}
    \label{fig:ramp_experiment}
\end{figure}

To anchor this calculation in a measurement, the drum was released from
rest on a shallow ramp of four parallel steel pipes
(Fig.~\ref{fig:ramp_experiment}). With four rails distributing the drum's weight rather than two, the
contact deformation at each rail is small enough that the effective
rolling radius coincides with $R_{\mathrm{out}}$ to within the
experimental precision quoted below, so no separate correction term is
introduced.Descent was filmed at 100~fps
($\pm 0.01$\,s resolution). From the travelled distance $S$ and
descent time $t$ the centre-of-mass acceleration is
\begin{equation}
    a = \frac{2S}{t^{2}}.
\end{equation}

Applying Newton's second law for translation and rotation,
$mg\sin\theta - T = ma$ and $TR = I\alpha$, together with
the no-slip condition $\alpha = a/R$, and eliminating the
friction force $T$, yields
\begin{equation}
    I = \left(\frac{mg\sin\theta}{2S}\,t^{2} - m\right)R^{2}.
    \label{eq:I_ramp}
\end{equation}

\begin{figure}[htbp]
    \centering
    \includegraphics[width=\columnwidth]{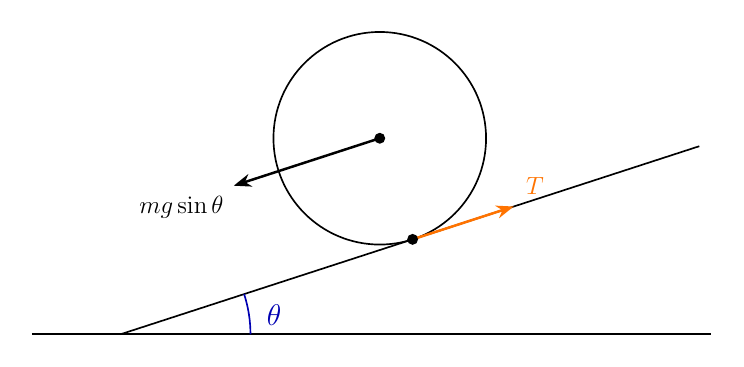}
    \caption{Free-body diagram for the drum on the rail ramp.
    $T$ is the friction force at the contact point and
    $mg\sin\theta$ is the gravitational component along the slope.}
    \label{fig:disk-incline}
\end{figure}

The measured parameters were $S = 3.900 \pm 0.001$\,m,
$\theta = 5.50^\circ$, $R = R_\mathrm{out} = 0.157 \pm 0.001$\,m, and
$t = 3.76 \pm 0.01$\,s. Evaluating Eq.~(\ref{eq:I_ramp}) with
standard quadrature error propagation gives
\begin{equation}
    I_\mathrm{exp} = 0.651 \pm 0.012~\mathrm{kg\,m^2},
\end{equation}
in excellent agreement with the geometric estimate of
$0.6507~\mathrm{kg\,m^2}$. We adopt the latter for all dynamometer
calculations.

\section{Data Acquisition}
\label{sec:daq}

The central measurement is the time interval between consecutive
revolutions. Capturing these at high fidelity is traditionally handled
by a dedicated Data Acquisition (DAQ) card costing several hundred euros. The key
instrumentation choice was to route an analogue optical signal into the
microphone input of a laptop, an approach demonstrated for precision
timing in several physics-education
contexts~\cite{Aguiar_2010,hassan2011soundcard}.

\subsection{Optical interrupter and community collaboration}

\begin{figure*}[t]
    \centering
    \includegraphics[width=\textwidth]{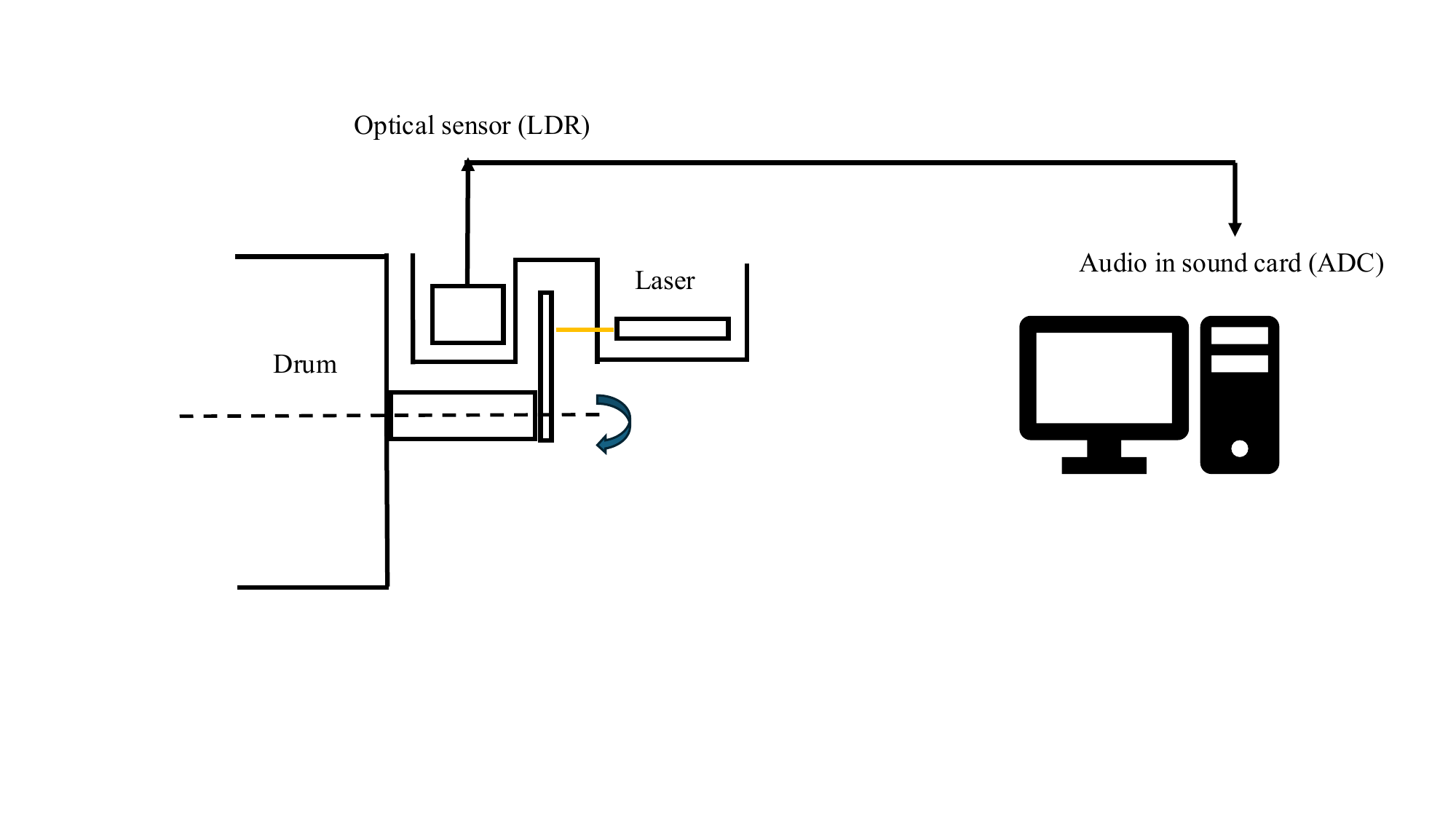}
    \caption{Block diagram of the data-acquisition chain. A radial
    blade on the drum shaft interrupts a collimated green laser beam;
    the light falling on the LDR is converted to a voltage pulse by
    the divider circuit and is recorded through the laptop microphone
    input at 44.1\,kHz.}
    \label{fig:daq_architecture}
\end{figure*}

The optical front-end was designed in collaboration with the director
of the regional Laboratory Centre of Physical Sciences (EKFE). A
single interceptor blade, welded radially to the main shaft, crosses
the beam of a green laser pointer once per revolution. A
cadmium-sulphide light-dependent resistor (LDR) placed opposite the
laser detects the interruption
(Fig.~\ref{fig:daq_architecture}).

\begin{figure}[htbp]
    \centering
    \includegraphics[width=0.95\columnwidth]{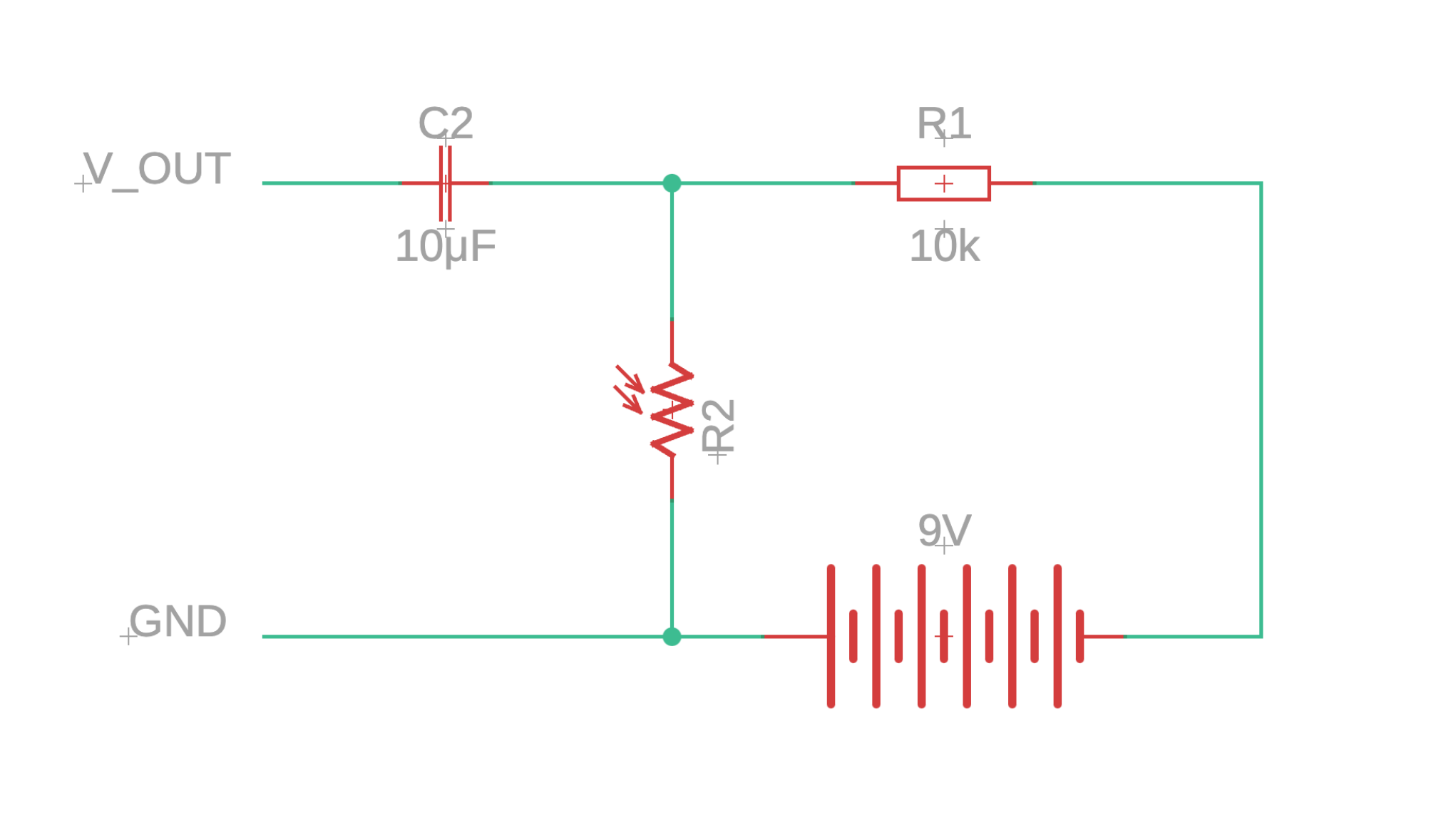}
    \caption{Optical-interrupter schematic. $R_1$ and $R_2$ form a
    voltage divider; the 10\,$\mu$F capacitor $C_2$ provides AC
    coupling so that only the pulse edges reach the sound-card input,
    isolating the laptop from the 9\,V supply.}
    \label{fig:circuit_schematic}
\end{figure}

The LDR ($R_2$) and a 10\,k$\Omega$ series resistor ($R_1$) form a
voltage divider driven by a 9\,V battery (Fig.~\ref{fig:circuit_schematic}):
\begin{equation}
    V_\mathrm{node} = 9\,\mathrm{V}\times\frac{R_2}{R_1+R_2}.
\end{equation}
When illuminated, $R_2$ drops below 100\,$\Omega$ and $V_\mathrm{node}
\rightarrow 0$; when the blade shadows the LDR, $R_2$ climbs into the
M$\Omega$ range and $V_\mathrm{node} \rightarrow 9$\,V. The resulting
waveform is a train of near-square voltage pulses at the drum rotation
frequency.

\subsection{Laptop sound card as an ADC}

A 10\,$\mu$F DC-blocking capacitor removes the DC component and,
with the sound-card input impedance
($Z_\mathrm{in} \sim 2$\,k$\Omega$), forms a high-pass filter at
$f_c = 1/(2\pi Z_\mathrm{in} C) \approx 8$\,Hz, protecting the
millivolt-rated codec while passing the pulse edges undistorted.
The sound card samples at $f_s = 44.1$\,kHz (16-bit), giving a
timing resolution of $1/f_s \approx 22.7~\mu$s, comfortably smaller
than any inter-pulse interval in the operating range. The recorded
WAV file is the raw data of the experiment.

\section{Signal Processing Pipeline}
\label{sec:signal}

The recorded audio trace (Fig.~\ref{fig:raw_data}) is processed in
MATLAB; an equivalent Python implementation and the original
\texttt{.m} script are provided as supplementary material, so the
pipeline is reproducible on entirely free software. There are four
steps.

\subsection{Noise reduction and peak picking}

A ten-point moving-average filter (\texttt{filtfilt}) is applied to
suppress electrical hum and high-frequency mechanical noise without
phase distortion. A dynamic threshold is then set at 50\,\% of the
maximum filtered amplitude, and the signal is subtracted from this
threshold. The resulting curve is interpolated by cubic splines and the timestamp of each pulse is then approximated with an appropriate algorithm \cite{Tsilioukas:2026repo}. This yields a sequence of timestamps $t_k$, one per revolution.

\subsection{Angular velocity}

Angular velocity is the straightforward finite difference
\begin{equation}
    \omega_k = \frac{2\pi}{t_{k+1}-t_k},
    \label{eq:omega}
\end{equation}
evaluated at the midpoint time
$\bar t_k = \tfrac{1}{2}(t_k+t_{k+1})$.

\subsection{Angular acceleration: the derivative problem}

A direct finite-difference estimate of $\alpha = d\omega/dt$ is
unusable in practice, because differentiation amplifies any noise
already present in $\omega$. We addressed this through a
polynomial smoothing step: $\omega(\bar t)$ is fitted by a polynomial
$p(t)$ whose analytical derivative $p'(t)$ gives a continuous
$\alpha(t)$.

The choice of polynomial order is itself a teaching point. A
tenth-order fit, tried first, proved unstable near the endpoints of
the sweep and produced non-physical oscillations, including negative
torque and negative power in the reconstructed curves. This is a
well-known artefact of high-order polynomial interpolation on
uniformly-spaced data. Reducing the order to three eliminated the
endpoint artefact and produced monotonic, physically sensible results;
the global shape of the sweep is well described by a cubic at the data
density available ($\sim$\,40 revolutions). Alternative smoothing
schemes such as the Savitzky--Golay
filter~\cite{savitzky1964smoothing} or smoothing splines were
discussed with the students as principled extensions. All results
shown in this paper use the third-order fit.

\subsection{Torque and power}

Torque and instantaneous mechanical power follow directly:
\begin{align}
    \tau(t) &= I\,\alpha(t), \\
    P(t)    &= \tau(t)\,\omega(t).
\end{align}
Plotting $\tau$ and $P$ as functions of $\omega$ rather than $t$ gives
the usual dynamometer presentation (Fig.~\ref{fig:final_curves}).

\section{Results}
\label{sec:results}

\subsection{Test vehicle and raw data}

\begin{figure}[htbp]
    \centering
    \includegraphics[width=\columnwidth]{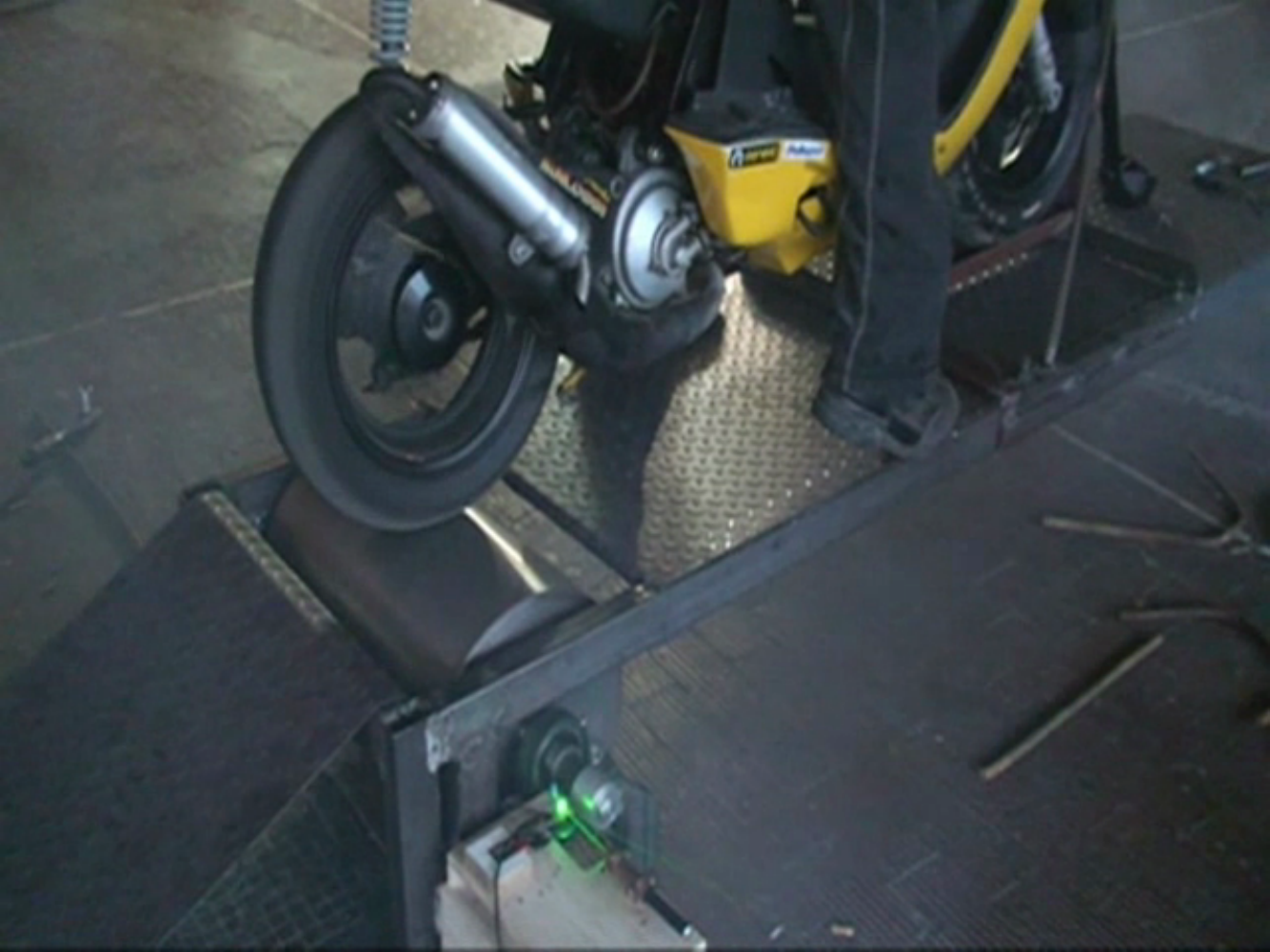}
    \caption{Wide-open-throttle test of the completed dynamometer
    with a tuned 50\,cc Piaggio NRG scooter. The green laser (lower
    right) illuminates the LDR through the interceptor blade on the
    drum shaft.}
    \label{fig:dyno_testing}
\end{figure}

The test vehicle was a tuned 50\,cc Piaggio NRG two-stroke scooter
with continuously variable transmission (CVT), chosen because it is
culturally ubiquitous among the EPAL student body and its CVT produces
a distinctive kinematic signature
(Fig.~\ref{fig:dyno_testing}).

\begin{figure}[htbp]
    \centering
    \includegraphics[width=\columnwidth]{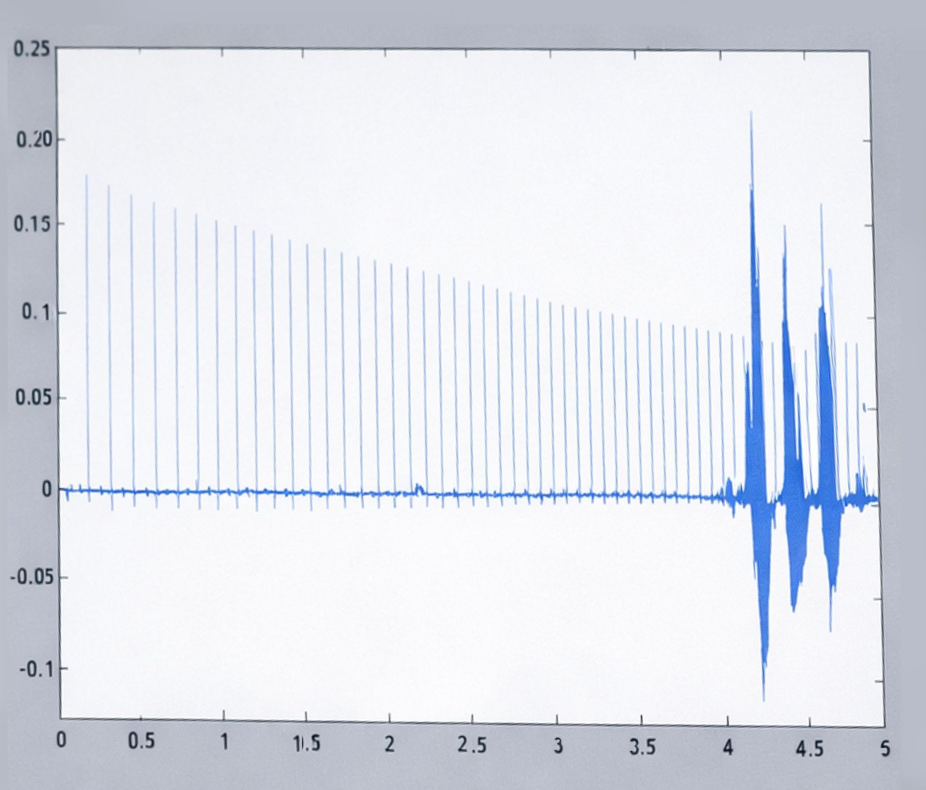}
    \caption{Raw sound-card recording of a Wide Open Throttle acceleration sweep.
    Decreasing inter-pulse spacing corresponds to increasing drum
    angular velocity. The amplitude envelope falls because the LDR,
    with a decay time of tens of milliseconds, is progressively less
    able to recover between pulses; beyond $t \approx 4.0$\,s the
    signal is no longer usable.}
    \label{fig:raw_data}
\end{figure}

A representative sweep (Fig.~\ref{fig:raw_data}) is clean from
$t = 0$ to $t \approx 4.0$\,s; beyond this the amplitude collapses
due to the sensor-bandwidth limit discussed in
Section~\ref{sec:bottleneck}.

\subsection{Torque and power curves}

\begin{figure}[htbp]
    \centering
    \includegraphics[width=\columnwidth]{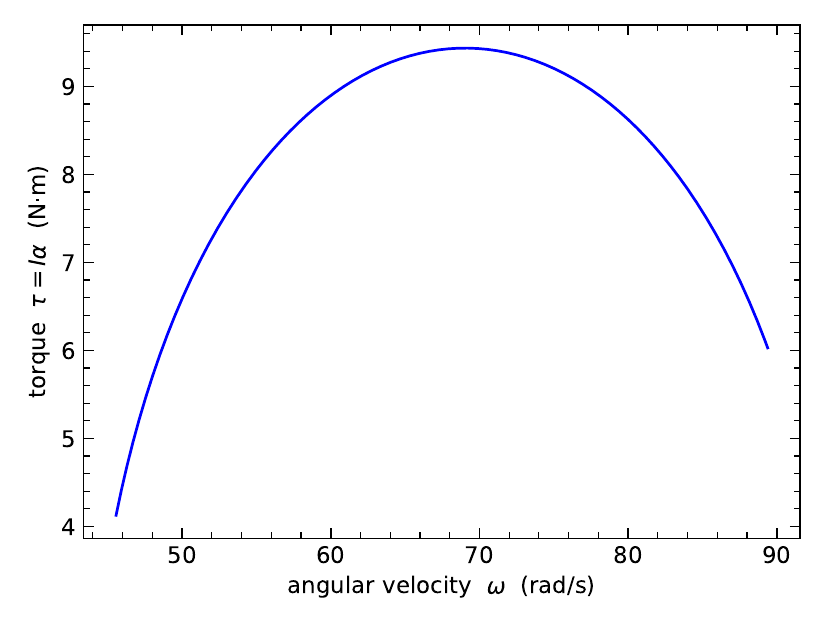}
    \vspace{0.4cm}
    \includegraphics[width=\columnwidth]{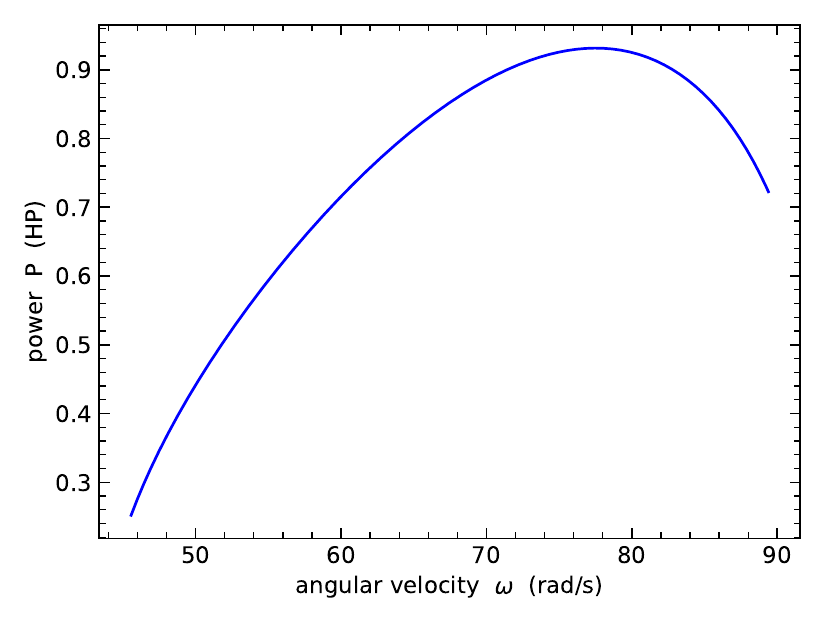}
    \caption{Reconstructed performance curves for the Piaggio NRG
    scooter. Top: torque $\tau = I\alpha$ versus drum angular
    velocity $\omega$. Bottom: mechanical power $P = \tau\omega$
    versus $\omega$. The shapes of the curves are consistent with
    the operation of a CVT, which tends to hold the engine near
    peak-power RPM while the drum accelerates, producing a broad
    plateau in $P$ and a falling branch in $\tau$.}
    \label{fig:final_curves}
\end{figure}

Before interpreting the reconstructed curves, it is useful to recall how
a scooter CVT works. Unlike a fixed ratio gearbox, it drives a belt
between two variable diameter pulleys: as engine RPM rises, centrifugal
weights in the drive pulley force the belt outward, continuously
adjusting the effective ratio so that the engine is held near the RPM at
which it produces maximum power, rather   than passing through a fixed
sequence of gears  \cite{Metwally2016DynamicPO}. Because the ratio is continuously adjusted, the
engine's own speed stays comparatively constant while the drum's angular
velocity $\omega$, the independent variable plotted in Fig.~\ref{fig:final_curves}, keeps
rising through the sweep. This is why the reconstructed curves
characteristically show torque falling and power remaining roughly
constant as $\omega$ increases: the engine is not losing capability; the
transmission is trading torque for rising drum speed at roughly constant
engine power output.
The pipeline of Section~\ref{sec:signal} yields the curves of
Fig.~\ref{fig:final_curves}. The torque rises to a maximum and then
falls while the power plateaus, qualitatively consistent with a CVT
holding the engine near peak-power RPM while the drum accelerates.
We stress that this is qualitative agreement with CVT theory \cite{Metwally2016DynamicPO}, not a
calibration against a reference dynamometer; absolute accuracy depends
on our estimate of $I$ and on the signal-processing approximations.

\section{The Sensor Bottleneck as a Teaching Tool}
\label{sec:bottleneck}

The signal collapse at $t \approx 4.0$\,s is not a sound-card
limitation but an LDR one. CdS photoresistors respond rapidly to
illumination onset (microsecond scale) but relax slowly when light is
removed; recovery times of 20--50\,ms are typical. As the drum
accelerates, the inter-pulse interval falls below the recovery time
and the LDR acts as an unintentional low-pass filter, blurring the
pulse train into a DC envelope.

This hardware failure became an instance of \emph{productive
failure}~\cite{kapur2008productive}. Students researched CdS
photoconductive properties, computed that at 10\,000\,RPM the pulse
period ($\Delta t \sim 6$\,ms) is well below the recovery time, and
proposed replacements: a photodiode with sub-microsecond response or a
Hall-effect sensor on a ferrous trigger wheel. Either replacement can be substituted directly into the
existing signal chain: like the LDR, both produce an analogue voltage
transition that can still be routed into the laptop's microphone input
and processed by the same pipeline described in Section~\ref{sec:signal}, so
resolving the bandwidth bottleneck does not by itself require
abandoning the sound card as ADC approach. A microcontroller based DAQ
(e.g.\ Arduino \cite{dausilio2012arduino}) remains a separate, independent
upgrade, offering higher timing resolution and native digital
triggering; this was beyond the 2010 scope of the project but remains
a promising next step.

\section{Pedagogical Observations}
\label{sec:pedagogy}

The project was not designed as an educational-research study and we
do not claim measured learning gains. We offer the following
observations in the spirit of a case report, as a possibly useful
data point for other VET practitioners. We interpret these observations through three distinct and complementary lenses rather than a single unifying theory: Bandura's self-efficacy theory, Lave and Wenger's situated learning, and Kapur's productive failure.

\textit{Role reversal and self-efficacy.} The welding phases placed
the students in the expert role; this inversion normalised the
subsequent physics-heavy phases. The mechanism is consistent with
Bandura's model~\cite{bandura1977self}: mastery experiences in
welding raised confidence to engage with unfamiliar physics.  

\textit{Community integration.} Collaboration with the local machine
shop, the EKFE director, and a neighbourhood carpenter situated the
project in a recognisable social network. In the language of Lave and
Wenger~\cite{lave1991situated}: learning is treated not as the transfer of decontextualised knowledge but as a byproduct of increasing participation in an authentic community of practice. In this sense the students moved from legitimate peripheral participants, assisting tradespeople, observing procedures, to active, trusted contributors, most visibly in the role-reversal of the welding phase, where they instructed the physics teacher directly.

\textit{Productive failure.} The sensor bottleneck forced an
unscripted engagement with datasheet-level semiconductor physics that
no planned lesson had delivered~\cite{kapur2008productive}.

\textit{Limitations of this account.} These observations are
retrospective and based on the instructor's field notes; no pre/post
conceptual inventories were administered. As such, we do not claim the three theoretical lenses above reduce to a single unifying framework; each is analysed through the theory that best accounts for it. A subsequent study, the
natural next paper, should replicate the build with a matched VET
cohort and measure learning gains against a standard
rotational-mechanics instrument such as the one developed by Mashood
and Singh~\cite{Mashood_2015}.

\section{Discussion}
\label{sec:discussion}

The apparatus reproduces, at relatively low cost, the core
measurement principle of a commercial inertia
dynamometer~\cite{martyr2012engine}: mapping a mechanical event (one
revolution of a calibrated flywheel) onto a digital timestamp with
sub-millisecond resolution. The sound-card-as-DAQ approach is not
original~\cite{Aguiar_2010,hassan2011soundcard}, but its
application to a student-built rotational dynamometer is, we believe,
novel and transferable.

The limitations are worth restating: the LDR imposes a hard upper
bound on resolvable angular velocity; the absolute torque scale depends
on the geometric estimate of $I$; and we report no
measured learning outcomes. A further limitation is analytical rather than mechanical: our situated-learning reading of the community-integration observations is offered as a retrospective interpretive lens, not as evidence of a designed apprenticeship trajectory. A fuller application of the framework would require tracing specific task hand-offs and shifts in responsibility over time, which a single-year, non-longitudinal case account cannot rigorously support. Each limitation is an entry point into a
richer successor project.   We describe the apparatus as
\emph{educational-grade}: accurate enough to demonstrate the qualitative
CVT signature. Nevertheless, the specific implementation cannot be used as a calibrated performance instrument. Hence, in
our view, the apparatus can be useful pedagogically. 
\\
We have presented a low-budget inertia dynamometer, built with adult
vocational students, that reproduces $\tau(\omega)$ and $P(\omega)$
curves for a small motorcycle across a limited RPM range. The physics
is the single equation $\tau = I\alpha$; the experimental craft lies
in every step between it and a recorded WAV file.

Natural extensions include: (i)~replacing the LDR with a photodiode or
Hall sensor; (ii)~adding a reference-torque calibration step;
(iii)~running a pre/post concept inventory to quantify learning gains;
and (iv)~porting the pipeline to free software (Octave/Python). The
supplementary material provides the MATLAB and Python source needed
to begin (iv) today.

\section*{Data and Code Availability}

The CAD design files (Fusion~360 and STEP formats), the MATLAB script
(\texttt{dynamometer.m}), and an equivalent Python implementation
(\texttt{dynamometer.py}) are available under an MIT licence at the publicly available repository \cite{Tsilioukas:2026repo}


\section*{Acknowledgements}

The author thanks the students of the 2010 Corfu EPAL evening physics
cohort, without whom no part of this project would exist; the
teachers, director and economic committee of the evening EPAL of Corfu for
supporting and financing the project; the local machine workshop Laertis Zorbas, for
the machining, welding, and dynamic-balancing work; the director of
the regional EKFE Panagiotis Mourouzis, for the electronics design collaboration; and the
carpenter Socrates Tryfonas. The authors have confirmed that any identifiable participants in this study have given their consent for publication

\bibliography{biblio}

\end{document}